\documentclass[preprint,12pt]{elsarticle}
\usepackage{graphicx}
\usepackage{amssymb}
\biboptions{square,comma,sort&compress}
\usepackage{url}

\journal{Chemical Physics Letters, accepted (2012)}

\begin{document}

\begin{frontmatter}

\title{NMR and NQR parameters of  ethanol crystal}

\author{M. Milinkovi\' c}

\author{G. Bilalbegovi\' c}

\address{Department of Physics, Faculty of Science, University of Zagreb,\\
Bijeni{\v c}ka 32, 10000 Zagreb, Croatia}

\date{\today}

\begin{abstract}
Electric field gradients and chemical shielding tensors
of the stable monoclinic crystal phase of ethanol are computed.
The projector-augmented wave (PAW) and gauge-including projector-augmented wave (GIPAW) models in the
periodic plane-wave density functional theory are used.
The crystal data from X-ray measurements, as well as
the structures where either all atomic, or only hydrogen atom
positions are optimized in the density functional theory are analyzed. These structural models are also studied
by including the semi-empirical Van der Waals correction to the density functional theory.
Infrared spectra of these five crystal models are calculated.
\end{abstract}


\end{frontmatter}

\section{Introduction}
\label{intro}

Organic crystals show many interesting phenomena and they are important in various applications \cite{Silinsh}.
The solid ethanol exists in several polymorphic phases \cite{Ramos}.
The stable crystal is monoclinic \cite{Jonsson}.
In addition, three glass phases exist: a plastic crystal, an orientational glass, and the true structural glass
\cite{Talon,Ramos}.
The glass phases have been investigated by X-ray diffraction \cite{Srinivasan},  Raman \cite{Srinivasan}, dielectric relaxation \cite{Benkhof}, calorimetric measurements \cite{Haida,Talon}, neutron scattering \cite{Talon},
EPR \cite{Kveder},  and NMR relaxation measurements \cite{Eguchi}.

The ethanol crystal belongs to the Pc space group and melts at 159 K \cite{Jonsson}. This crystal consists of zig-zag molecular chains which are connected by hydrogen bonding. Two types of molecules are present in each chain: one is with a C-C-O-H torsion angle of -63$^{\circ}$, and other is a conformation with a torsion angle of 179$^{\circ}$. It is not easy to grow good quality monoclinic crystals
because of complex thermodynamic properties of the solid ethanol.
The ethanol molecule  is known as one of the first systems for which NMR measurements in organic solvents were carried out \cite{Arnold}.
The temperature dependence of the relaxation time in various phases of ethanol, including the
monoclinic solid, have been studied by NMR \cite{Eguchi}.
In spite of intensive nuclear resonance studies of various organic materials and a substantial improvement of experimental techniques, we have not been able to find any report of measurements for chemical shifts and quadrupolar parameters
in the monoclinic solid phase of ethanol.

Advances in magnetic resonance experiments have been followed by a development of the first principle calculations of corresponding parameters. These theoretical approaches help in analysis of often complicated NMR spectra  by connected them to the underlying microscopic structure of materials.
Recently, the density functional theory (DFT) based
gauge-including projector-augmented wave (GIPAW) method  has been developed \cite{Mauri,Harris,Yates,Charpentier}.
In a comparison with older quantum chemistry techniques, which use the cluster approximation,
GIPAW is particularly suitable for computations of the magnetic response of crystals.
This method works with periodic boundary conditions and therefore correctly  describes extended periodic systems, for example  lattice effects and
non-covalent interactions often present in molecular crystals.
In addition, GIPAW is an extension of the projector augmented wave (PAW) method \cite{Blochl}. PAW and GIPAW
use special pseudopotentials which, by reconstructing all-electron density close to atomic nuclei,
correctly account for electrons  in this region. PAW method has been successfully applied
in calculations of nuclear quadrupolar parameters in the density functional approach
\cite{Petrilli,Profeta}.

It has been found that the GIPAW method produces adequate results for
a wide range of organic solids
\cite{Harris,Payne,Johnston,Kibalchenko,Ceresoli}.
In this study we present a DFT plane wave calculation of the ethanol crystal. We calculate corresponding  hydrogen, carbon, and oxygen
chemical shieldings in the GIPAW model.  Quadrupolar coupling constants and asymmetry parameters for $^2H$ and $^{17}O$
are computed using the PAW method.

In contrast to NMR and NQR parameters, measurements of infrared spectra of the ethanol crystal have been carried out 
\cite{Mikawa,Rozenberg}.
Infrared spectra of the ethanol crystal, as well as of thin and thick solid ethanol films deposited on several substrates, have also been measured because of applications in astrophysics \cite{Boudin,Burke}.
Using density functional perturbation theory methods \cite{Baroni}, we calculate infrared spectra of the ethanol crystal
and compare these results with experiments.

\section{Computational methods}
\label{methods}

NMR and NQR parameters of the monoclinic crystal of ethanol are computed
using the GIPAW  and PAW
models \cite{Mauri,Harris,Yates,Charpentier} implemented in the Quantum ESPRESSO  suite of computer codes for electronic-structure calculations and materials modeling at the nanoscale \cite{Giannozzi}.
The crystal structure and atomic positions are taken from the Cambridge Structural Database \cite{Allen} where data measured  by the X-ray diffraction at 87 K \cite{Jonsson} are available.
Electronic ground state and NMR calculations  using directly experimental atomic positions have been carried out
in similar DFT studies for crystals whose structures are well known and used \cite{Harris}. We take this approach here in our first crystal model
(labeled as $a$ in the results), but also studied several other geometries. In the second model ($b$), we take the same data from the Cambridge Structural Database, but then we optimize all
atomic positions in DFT within the fixed unit cell. A such approach is usually taken for older experimental data obtained by X-ray diffraction measurements in which positions of atoms were often determined with large errors.
In the model $c$, when relaxing all atomic positions
we added the semi-empirical Van der Waals correction at the DFT-D2 level
\cite{Grimme,Barone}. This correction has been tested for large benchmark sets of molecules, clusters, surfaces, and crystals
where the Van der Waals interaction is important, and good agreement with available experimental results have been found.
It is known that the largest errors in older X-ray measurements of organic materials exist for H atom positions \cite{Harris}.
Therefore, we also study the models $d$, where only hydrogen atoms are relaxed in DFT, and $e$ where only hydrogen atoms are relaxed in DFT corrected
by the Van der Waals interaction.

The generalized gradient approximation (GGA) of Perdew, Burke, and Ernzerhof \cite{Perdew}, and
norm-conserving GIPAW pseudopotentials prepared by A.P. Seitsonen \cite{Seitsonen} are applied.
The cutoff energy of 90 Ry is used. The Brillouin zone is sampled with the 12$\times$10$\times$8 Monkhorst-Pack set of k points \cite{Pack}.
We found that the f-sum rule is fulfilled under these conditions.
As reference systems we study the isolated tetramethylsilane (TMS) and
ethanol molecules positioned in a box of 40 Bohr. For these systems we also use the cutoff of 90 Ry.''

The external magnetic field ${\vec B}$ and an effective field ${\vec B_{in}({\vec r})}$ in a material are connected by
\begin{equation}
{\vec B_{in}({\vec r})} = -{\hat \sigma} ({\vec r}) {\vec B},
\label{eq1}
\end{equation}
where ${\hat \sigma} ({\vec r})$ is the nuclear shielding tensor.
Chemical shifts are often calculated from
\begin{equation}
\delta({\vec r})=\sigma_{ref} - \sigma ({\vec r}),
\label{eq2}
\end{equation}
where $\sigma_{ref}$ is defined for a chosen reference nucleus, and $\sigma ({\vec r})=1/3Tr[{\hat \sigma} ({\vec r})]$.
The parameters typically used to describe quadrupolar interactions are the coupling constant $C_Q$ and the asymmetry parameter
$\eta$.
The quadrupolar coupling constant is
\begin{equation}
C_Q = \frac{eV_{zz}Q}{h},
\label{eq3}
\end{equation}
where $e$ is the  electron charge, $V_{zz}$ is the $z$ principal component of the EFG  tensor,
$Q$ is the nuclear quadrupolar moment, and
$h$ is the Planck constant.
The asymmetry parameter is
\begin{equation}
\eta_Q = \frac{V_{xx}-V_{yy}}{V_{zz}},
\label{eq4}
\end{equation}
where the components of the EFG tensor satisfy the condition
$|V_{zz}|> |V_{yy}|>|V_{xx}|$.
Quadrupolar moments for nuclei we study are:
$Q(^2H) =2.860\times 10^{-31}$ m$^2$ and $Q(^{17}O)= -25.58 \times 10^{-31}$ m$^2$.
\cite{Pyykko}.
Using  these values of nuclear quadrupolar moments and Eqs. (1)-(4) we calculate NMR and NQR parameters of the ethanol crystal in the GIPAW and PAW  models.

Infrared spectra of five models of the ethanol crystal are calculated 
to study the stability  of lattices and to compare results with experiments. 
The methods based on a density functional perturbation theory \cite{Baroni} and the Quantum ESPRESSO 
package \cite{Giannozzi} are applied. The same GIPAW pseudopotentials as in NMR/NQR calculations are used.
We have been increasing thresholds for self-consistency in ground state and phonon calculations (to $10^{-12}$ Ry and $10^{-20}$, respectively), as well as the cutoff to 110 Ry, in order to check the ability of crystal models to get rid of negative frequencies of normal modes.

\section{Results and Discussion}
\label{results}

Table 1 shows bond lengths for different strategies in the optimization procedure.
The average value is taken where more bonds of the same type exist.
The O-C and C-C  bonds are with similar lengths
in all models of atomic positions and agree at the 0.01 \AA\, level.
Differences in bond lengths with H atoms are larger.
The C-H, O-H, and H bonds in methyl groups  increase in DFT optimized models in a comparison to the experimental data.
The lengths of hydrogen bonds decrease in DFT-optimized models.
It is possible to explain differences between experimental  and optimized atomic positions by always present problems in DFT (for example, approximations in  exchange functionals and pseudopotentials), but also by  the fact that
X-ray measurements \cite{Jonsson} are rather old. Therefore, as in other similar studies, positions of atoms are sometimes determined with
large errors. In addition, experiments were carried out at 87K, whereas DFT calculations neglect thermal effects.
Differences between DFT-optimized bond lengths without and with the van der Walls corrections
(i.e., $b$ vs $c$ and $d$ vs $e$ sets) are nonexistent or small. The largest change is at the
0.01 \AA\, level when all atomic positions are optimized ($b$ vs $c$ ), and at 0.001 \AA\, when only H atoms are relaxed ($d$ vs $e$).
However, it is known that calculations of NMR parameters are sensitive even to small changes of atomic positions.

The chemical shieldings are shown in Table 2, and the corresponding atomic labeling is explained in Fig. 1.
The results show that the largest shieldings exist for hydrogen atoms in the OH group, then for ones from the methyl group, and the smallest are for
H in the CH$_2$ group.
Hydrogen atoms in the same methyl group have slightly different shieldings.
Carbon atoms from the methyl group exhibit larger shieldings than C in the CH$_2$ group.
Shieldings in DFT-optimized models  are much smaller then ones calculated for positions from the X-ray measurements.
It is important to point out that our calculations  do not consider temperature effects.
NMR and NQR calculations are carried out for static molecules in the crystal lattice and for fixed atomic positions.

The results of density functional theory calculations are the components of the nuclear shielding tensor.
Different approaches are taken in the literature to compare
these results with experiments. Experimental results for the ethanol crystal are not available. Therefore, we are not able to fix the scale
using the fitting to the experimental data. Instead,
calculated here shieldings in isolated molecules of
TMS (for hydrogen and carbon), and ethanol (for oxygen)
are taken as reference values: $\sigma (H)= 30.4$ ppm,
$\sigma (C)= 178.2$ ppm,
$\sigma (O)= 258.6$ ppm.
We did not use average values for $\sigma (C)$ and  $\sigma (H)$ calculated in the ethanol molecule because of the fact that
these atoms exist (in a molecule, as well as in the crystal)
in different geometrical and chemical environments. In contrast, shieldings are the same
for all four carbon and all twelve hydrogen atoms in TMS.
Because of the lack of the experimental fixing scale, and to benefit possible future
experiments, we report both shieldings (Table 2) and shifts (Table 3).

\begin{table}
\centering{
\caption{Bond lengths (\AA) in the ethanol crystal, $a$: X-ray atomic positions \cite{Jonsson}, $b$: all atomic positions are optimized in DFT,
$c$: all atomic positions are optimized in DFT corrected by the van der Waals interaction,
$d$: positions of hydrogen atoms are optimized in DFT, $e$: positions of hydrogen atoms are optimized in DFT corrected by the van der Waals interaction as explained in the text.}
\begin{tabular}{c c c c c c}
\hline
Bond & $a$ & $b$ & $c$ & $d$ & $e$\\
\hline
O-C  &   1.427             &    1.428                 &  1.429      &  1.427    &  1.427    \\
C-C  &   1.506          &   1.509               &  1.512        & 1.506 &  1.506     \\
C-H  &   0.974           &  1.107                 & 1.108     & 1.105 &  1.106         \\
O-H  &   0.822         &    0.982               &  0.982              &0.982 & 0.982 \\
Hydrogen O-H  &  1.927           &  1.880                 & 1.868     &1.730 &  1.730       \\
Methyl C-H &  0.987            &  1.098                &  1.099    &1.098 &  1.099         \\
\hline
\end{tabular}
\label{table1}
}
\end{table}

\begin{figure}
\centering
\includegraphics[scale=0.5]{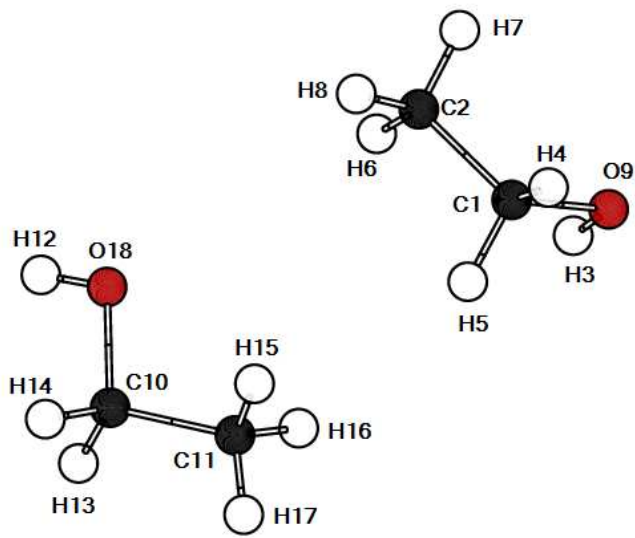}
\caption{
The labeling scheme for C, H, and O atoms used in Tables 2-5.}
\label{fig1}
\end{figure}

\begin{table}
\centering{
\caption{ The NMR shieldings $\sigma $ (ppm) in the ethanol crystal, $a$: atomic positions from X-ray measurements \cite{Jonsson},
$b$: all positions are optimized in DFT,
$c$: all positions are optimized in DFT corrected by the van der Waals interaction,
$d$: positions of hydrogen atoms are optimized in DFT, $e$: positions of hydrogen atoms are optimized in DFT corrected by the van der Waals interaction.
Atom numbers are shown in Fig. 1.}
\begin{tabular}{c c c c c c}
\hline
Nucleus & $\sigma$($a$) & $\sigma$($b$)  & $\sigma$($c$) & $\sigma$($d$) & $\sigma$($e$)\\
\hline
C1 (CH$_2$ group) & 130.2 & 109.2&108.8 & 110.7& 110.2\\
C10 (CH$_2$ group) & 135.8 & 111.9&111.5 &111.7 &111.2\\

C2  (methyl group)& 175.6 & 153.9 & 152.6& 155.1& 154.6\\
C11  (methyl group)& 188.0& 156.9 & 155.8& 158.1& 157.6\\		
\hline	
H3 (OH group) &35.6  & 30.8 & 30.7&31.4 & 31.0\\
H12 (OH group) & 36.3 & 26.6 & 26.3&  25.7& 25.4\\

H6 (methyl group) & 31.4 & 29.5 & 29.2&30.4 & 29.9\\
H7 (methyl group)& 32.6 &  29.2 & 28.9 & 29.7& 29.4\\
H8 (methyl group)& 33.7 &  29.6 & 29.3&30.0 & 29.7\\
H15 (methyl group) & 34.9 & 30.0&29.7 &30.7 & 30.3\\
H16 (methyl group)& 33.3  & 30.0 &29.8 & 30.7 & 30.4\\
H17 (methyl group)& 34.7 & 29.7 &  29.4& 29.6& 29.1\\

H4  (CH$_2$ group) & 31.4 &27.1 &26.9 &27.7 &  27.4\\
H5 (CH$_2$ group)& 28.6  & 26.3 &26.0 & 26.8 & 26.5\\
H13 (CH$_2$ group) & 30.4 & 26.5 & 26.3 &  27.3& 26.9\\
H14 (CH$_2$ group)& 30.4 & 27.0 & 26.8& 27.6& 27.2\\
\hline
O9  & 314.9 & 255.3 &254.5 & 253.0& 252.7\\
O18  & 335.3 & 274.5 &273.5 &  268.6& 267.7\\
\hline
\end{tabular}
\label{table2}
}
\end{table}

\begin{table}
\centering{
\caption{ The NMR chemical shifts $\delta $ (ppm) in the ethanol crystal, $a$: X-ray atomic positions \cite{Jonsson}, $b$: all atomic positions are optimized in DFT,
$c$: all atomic positions are optimized in DFT corrected by the van der Waals interaction,
$d$: positions of hydrogen atoms are optimized in DFT, $e$: positions of hydrogen atoms are optimized in DFT corrected by the van der Waals interaction.
The shieldings in TMS and ethanol molecules (calculated here under similar
conditions as for the crystal) are taken as reference values.
Atom numbers are shown in Fig. 1.}
\begin{tabular}{c c c c c c}
\hline
Nucleus & $\delta$($a$) & $\delta$($b$)  & $\delta$($c$) & $\delta$($d$) & $\delta$($e$)\\
\hline
C1 (CH$_2$ group) & 48.0 & 69.0 & 69.4 & 67.5& 68.0\\
C10 (CH$_2$ group) & 42.4 & 66.3& 66.7& 66.5& 67.0\\

C2 (methyl group) & 2.6 & 24.3 & 25.6& 23.1& 23.6\\
C11 (methyl group) & -9.8& 21.3 & 22.4& 20.1& 20.6\\		
\hline	
H3 (OH group) &-5.2  & -0.4 &-0.3 & -1.0& -0.6\\
H12 (OH group) &-6.0 & 3.8 & 4.0&4.7 &5.0\\

H6 (methyl group) & -1.0 & 0.9 & 1.2& 0.0& 0.5\\
H7 (methyl group)& -2.2 &  1.2 &1.5 & 0.7& 1.0\\
H8 (methyl group)& -3.3 & 0.8  & 1.1& 0.4& 0.7\\
H15 (methyl group) & -4.5 &0.4  &0.7 & -0.3& 0.1\\
H16 (methyl group)& -2.9  &0.4 & 0.6& -0.3& 0.0\\
H17 (methyl group)& -4.3 & 0.7 &1.0 &0.8 &1.3\\

H4 (CH$_2$ group) & -1.0 & 3.3&3.5 & 2.7& 3.0\\
H5 (CH$_2$ group)& 1.8  & 4.1 & 4.4&3.6 &3.9\\
H13 (CH$_2$ group) & 0.0  & 3.9 & 4.1& 3.1&3.5\\
H14 (CH$_2$ group)& 0.0 &3.4 & 3.6 &2.8 &3.2\\
\hline
O9  & -56.3  & 3.3 &4.1 &5.6 &5.9\\
O18  &-76.7  & -15.9 &-14.9 &-10.0 &-9.1\\
\hline
\end{tabular}
\label{table3}
}
\end{table}

Quadrupolar parameters are shown in Table 4 and Table 5. Although existing NQR experiments do not provide signs of quadrupolar coupling constants we, in principle, show them for calculated quadrupolar parameters. However, all calculated quadrupolar parameters are positive.
Coupling constants for H decrease, and for O atoms increase  in DFT optimized models. Results for all coupling constants and asymmetry parameters in the DFT-optimized models $b-e$ are the same or similar to each other. The largest differences between the model $a$ and DFT-optimized models $b-e$ exist for coupling constants of hydrogen atoms in the OH group.

\begin{table}
\centering{
\caption{Quadrupolar coupling constants $C_Q$ (in MHz) of the ethanol crystal, $a$: X-ray atomic positions \cite{Jonsson}, $b$: all atomic positions are optimized in DFT,
$c$: all atomic positions are optimized in DFT corrected by the van der Waals interaction,
$d$: positions of hydrogen atoms are optimized in DFT, $e$: positions of hydrogen atoms are optimized in DFT corrected by the van der Waals interaction.
Atom numbers are shown in Fig. 1.}
\begin{tabular}{c c c c c c}
\hline
Nucleus & $C_Q(a)$ & $C_Q(b)$ & $C_Q(c)$ & $C_Q(d)$ & $C_Q(e)$\\
\hline	
H3  (OH group)& 0.60    & 0.26 &  0.26 &0.26 &0.26\\
H12  (OH group)& 0.89  & 0.22& 0.22  & 0.22 &0.21\\

H6 (methyl group) &0.22  &0.18 &0.18  &0.18  &0.18\\		
H7 (methyl group) & 0.30 &0.17 & 0.16 & 0.17&0.17\\
H8  (methyl group)& 0.39 & 0.18 & 0.18 & 0.18 &0.18\\
H15 (methyl group) & 0.42 & 0.18 & 0.18  &0.18 &0.18\\
H16 (methyl group) & 0.29& 0.18& 0.18 &0.18 &0.18\\
H17 (methyl group) & 0.44  &0.19 & 0.18 &0.18 &0.18\\

H4 (CH$_2$ group) & 0.46  &0.18  &  0.17  &  0.18&0.17\\
H5  (CH$_2$ group)& 0.25 & 0.15 & 0.15 &0.16 &0.15\\
H13 (CH$_2$ group) & 0.38 & 0.16 & 0.16  &0.17 &0.17\\
H14 (CH$_2$ group)& 0.36 & 0.17& 0.17 & 0.17  &0.17\\
\hline
O9 &10.92   &11.47 & 11.44  &11.50 &11.48\\
O18  &10.47  &10.92 & 10.90 & 10.56 &10.57\\
\hline
\end{tabular}
\label{table4}
}
\end{table}

\begin{table}
\centering{
\caption{Quadrupolar asymmetry parameters $\eta _Q$ of the ethanol crystal,  $a$: X-ray atomic positions \cite{Jonsson}, $b$: all atomic positions are optimized in DFT,
$c$: all atomic positions are optimized in DFT corrected by the van der Waals interaction,
$d$: positions of hydrogen atoms are optimized in DFT, $e$: positions of hydrogen atoms are optimized in DFT corrected by the van der Waals interaction.
Atom numbers are shown in Fig. 1.}
\begin{tabular}{c c c c c c}
\hline
Nucleus & $\eta _Q(a)$ & $\eta _Q(b)$ & $\eta _Q(c)$ & $\eta _Q(d)$ & $\eta _Q(e)$\\	
\hline	
H3  (OH group)& 0.10  & 0.14 & 0.14 &0.15 & 0.14\\
H12  (OH group)& 0.08 & 0.16&0.16  & 0.16& 0.16\\

H6 (methyl group) &0.04  & 0.03&0.03  &0.04 &0.04\\		
H7 (methyl group) &0.02  & 0.02& 0.02 &0.03 &0.03\\
H8  (methyl group)& 0.02 & 0.02& 0.02 &0.03 &0.03 \\
H15 (methyl group) & 0.02 &0.02 & 0.02 &0.03 &0.03\\
H16 (methyl group) &0.03 & 0.02& 0.02 & 0.03& 0.03\\
H17 (methyl group) & 0.02 &0.03 & 0.03 & 0.03&0.03\\

H4 (CH$_2$ group) & 0.05  &0.08 &0.08 &0.08 &0.08 \\
H5  (CH$_2$ group)& 0.06  &0.07 &0.07  &0.08 &0.08\\
H13 (CH$_2$ group) & 0.05 &0.06 & 0.06 &0.05 &0.05\\
H14 (CH$_2$ group)& 0.05 &0.06 & 0.06 &  0.05&0.05\\
\hline
O9 &0.75  &0.81 & 0.81 & 0.82&0.81\\
O18  & 0.98  & 0.88& 0.88 & 0.92&0.91\\
\hline
\end{tabular}
\label{table5}
}
\end{table}

\begin{figure}
  \begin{center}
    \begin{tabular}{c}
       \resizebox{100mm}{!}{\includegraphics{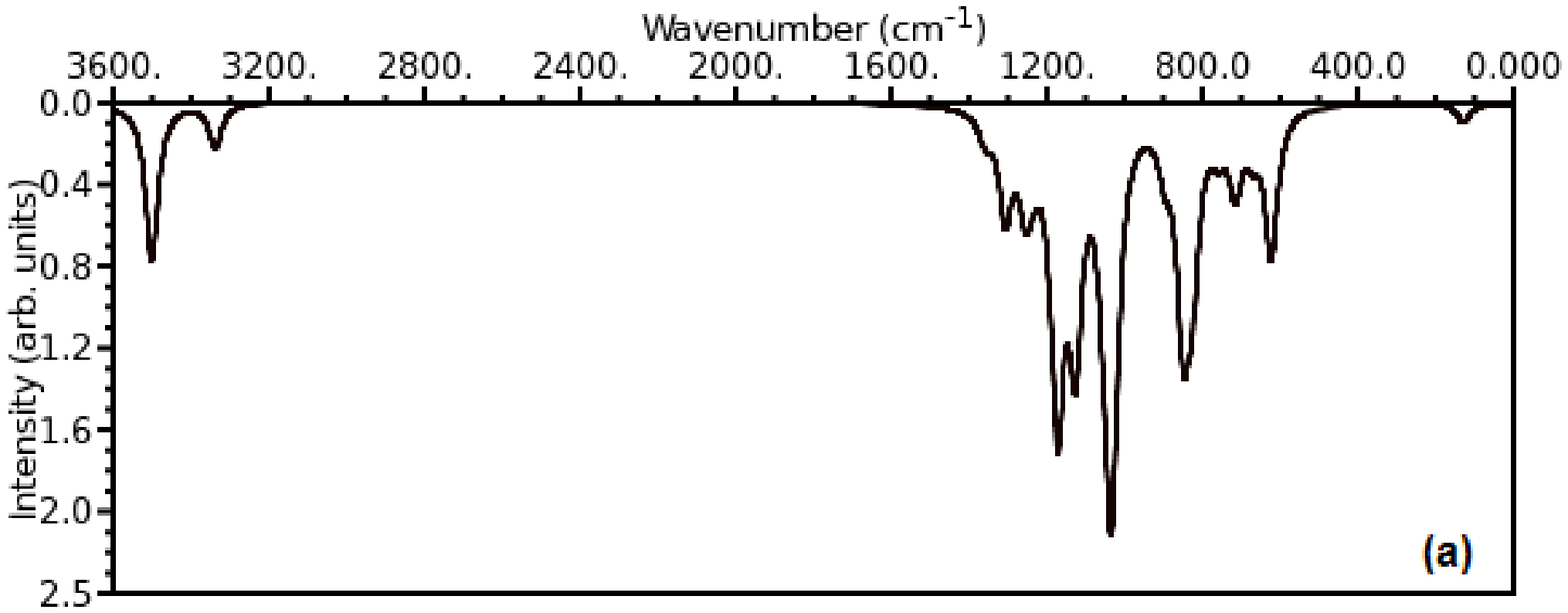}} \\
       \resizebox{100mm}{!}{\includegraphics{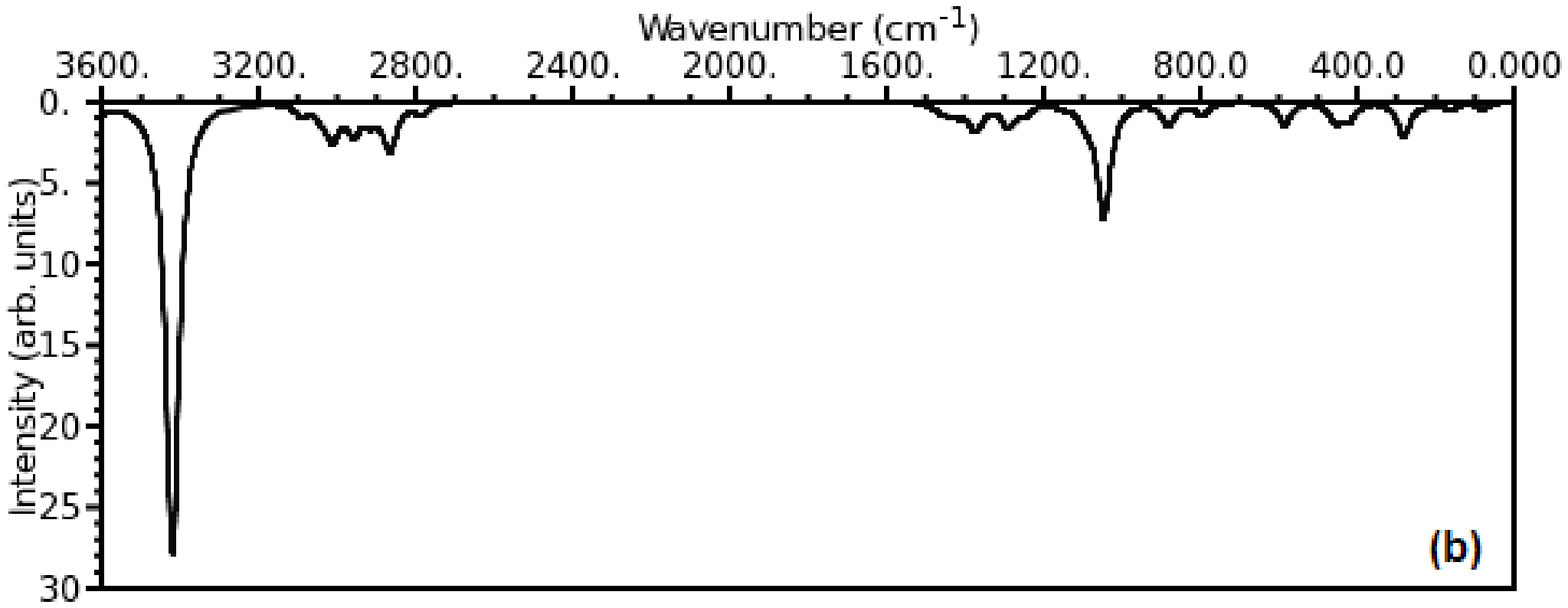}} \\
       \resizebox{100mm}{!}{\includegraphics{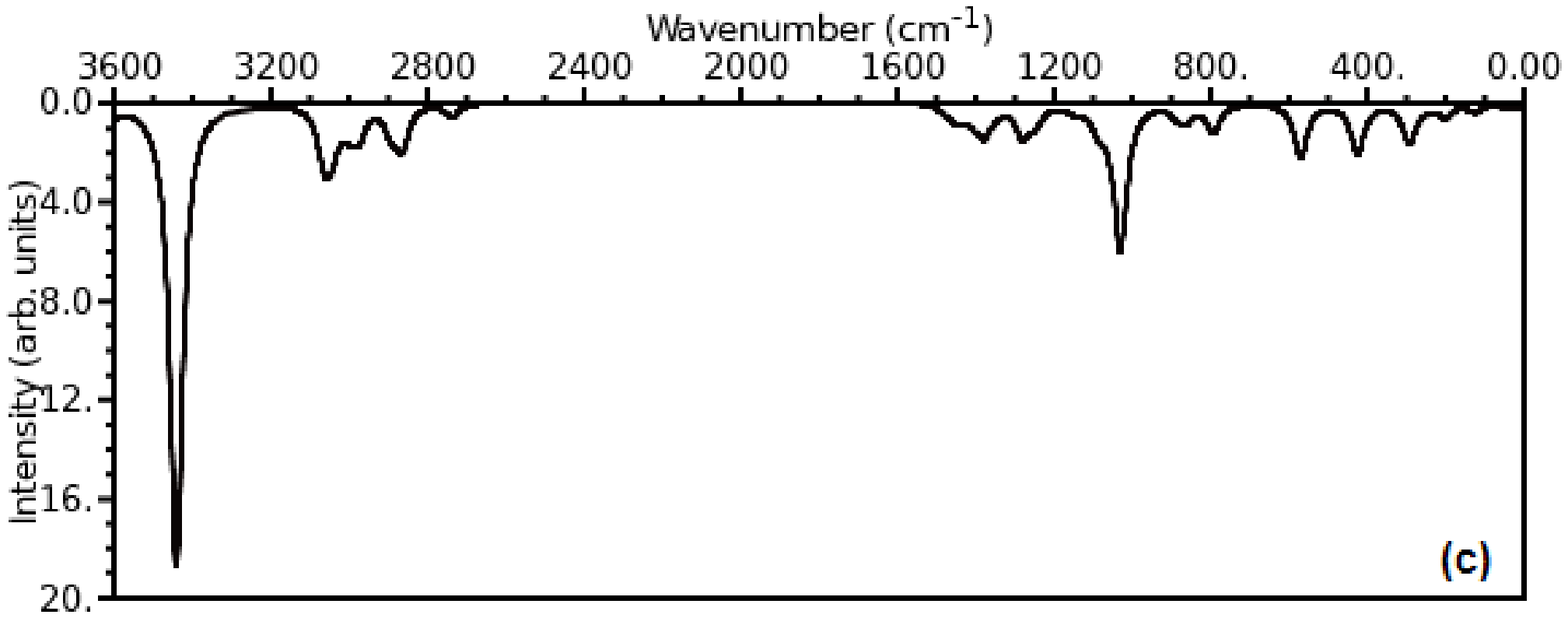}} \\
    \end{tabular}
    \caption{Infrared spectra of ethanol crystal calculated by density functional perturbation theory methods, 
(a) model  $a$: X-ray atomic positions \cite{Jonsson}, (b) model $b$: all atomic positions are optimized in DFT, (c) model $d$: positions of hydrogen atoms are optimized in DFT. Model $a$ is unstable (see text).}
    \label{fig2}
  \end{center}
\end{figure}

Infrared spectra for crystal structures $a$, $b$ and $d$ are shown in Fig. 2. 
We have calculated infrared spectra for all five models. However, the results show
that small changes in coordinates induced by the van der Waals corrections (models $c$ and $e$) do not produce substantial changes in infrared spectra.
The model $a$ shows many ($\sim 28\%$) negative vibrational frequencies. In contrast to the models $b$, $c$, $d$, and $e$, the number of negative frequencies in the
model $a$ does not decrease with changes of  parameters typically used in studies of vibrational properties of materials by density functional perturbation theory methods \cite{Baroni,Giannozzi}. 
These parameters are accurate self-consistency thresholds in the ground state and phonon calculations, as well as the larger energy cutoffs.
The persistence of many negative frequencies only in the model $a$, indicates its inherent instability. Therefore, as for the bond lengths,
the calculation of vibrational properties of five ethanol crystal models shows that new measurements of atomic positions in
the ethanol crystal are necessary. 
Spectra in Fig. 2 (b) and  Fig. 2 (c) show lines in two intervals of wavenumbers: (500-1500) cm$^{-1}$ and (2800-3500) cm$^{-1}$,
where IR peaks have also been measured \cite{Mikawa,Rozenberg}.  
In contrast, the spectrum of model $a$ (Fig. 2 (a)) does not have lines in the interval (2800-3300) cm$^{-1}$. 
Several far infrared lines in Fig. 2 (b) and  Fig. 2 (c) also exist, in good agreement with measured data: (78-279) cm$^{-1}$
\cite{Mikawa}.

\section{Conclusions}
\label{concl}
Using the GIPAW and PAW DFT methods we calculate NMR chemical shielding tensors and NQR parameters of the ethanol crystal.
To the best of our knowledge,
corresponding experimental NMR and NQR data are not available.
We consider several sets of atomic positions: experimental from X-ray diffraction measurements,
optimized in the GGA DFT model, and optimized in the GGA DFT model corrected by
the van der Waals interaction. Situations where only hydrogen atoms are allowed to move, or where all atoms move,  are studied in the optimization procedure.
Differences in atom positions between DFT-optimized models and rather old X ray measurements
are found. It is also calculated that shieldings substantially decrease when all atom positions
are relaxed in density functional models.
Results for infrared spectra of crystal models where positions of either hydrogen, or all atoms are optimized, show good agreement with
measured data. In contrast, the crystal model where all coordinates are taken from X ray measurements, in our calculations shows instability, as well as
the absence of intervals of measured IR lines.
New measurements of atomic positions in the ethanol crystal are desirable.
We hope that our results will be helpful for analysis and assignments in future measurements of NMR and NQR  parameters for the ethanol crystal.

\section*{Acknowledgement}
This work has been done within the HR--MZOS project 119--1191458--1011.
The authors are grateful to the University of Zagreb Computing Centre SRCE for a computer time and their help, as well
as to Marina Kveder and Miroslav Po\v zek for discussions.

\end{document}